\documentclass[aps,onecolumn,superscriptaddress]{revtex4-2}

\usepackage[T1]{fontenc}
\usepackage{charter}

\usepackage{amsmath,amssymb,amsfonts,color,graphicx,tabularx}

\usepackage[unicode=true,colorlinks=true]{hyperref}

\hypersetup{linkcolor=blue,citecolor=blue,urlcolor=blue}

\begin{document}

\title{Exciton condensation of composite fermions in double layer quantum Hall systems}

\author{Xiang-Jian Hou}
\affiliation{School of Physics and Wuhan National High Magnetic Field Center, Huazhong University of Science and Technology, Wuhan 430074, China}

\author{Lei Wang}
\affiliation{National Laboratory of Solid-State Microstructures, Collaborative Innovation Center of Advanced Microstructures, School of Physics, Nanjing University, Nanjing, 210093, China}

\author{Ying-Hai Wu}
\email{yinghaiwu88@hust.edu.cn}
\affiliation{School of Physics and Wuhan National High Magnetic Field Center, Huazhong University of Science and Technology, Wuhan 430074, China}

\begin{abstract}
We study fractional quantum Hall states in double layer systems that can be interpreted as exciton condensates of composite fermions. An electron in one layer is dressed by two fluxes from the same layer and two fluxes from the other layer to become composite fermions that form effective Landau levels. It is found that two types of composite fermion exciton condensates could occur. In the first type ones, all effective levels are partially occupied and excitonic correlations are present between composite fermions in the same effective level. In the second type ones, composite fermions in the topmost effective levels of the two layers form exciton condensate whereas those in lower effective levels are independent. The electric transport signatures of these states are analyzed. We demonstrate using numerical calculations that some composite fermion exciton condensates can be realized in microscopic models that are relevant for graphene and transition metal dichalcogenides. For a fixed total filling factor, an exciton condensate may only be realized when the electron densities in the two layers belong to a certain range. It is possible that two types of states appear at the same total filling factor in different ranges. These results shed light on recent experimental observations and also suggest some promising future directions. 
\end{abstract}

\maketitle

\section{Introduction}
\label{intro}

As a fundamental constituent of matter, electrons carry negative charge. In solid state systems, collective motion of electrons leads to a large variety of phenomena, but some of them appear to indicate the presence of mobile objects with positive charge. This fact can be described conveniently using holes, which are vacancies left by electrons in otherwise fully occupied bands. When a system hosts both electrons and holes with sufficiently strong correlation, they may form bound states that are called excitons, which exhibit collective phenomena such as Bose-Einstein condensation (BEC)~\cite{Blatt1962,Jerome1967,Kohn1970}. While this is a tantalizing prospect, experimental demonstration turns out to be extremely challenging. It is somewhat surprising that convincing observations were first reported in two-dimensional electron gases in the quantum Hall regime~\cite{Eisenstein2004,Eisenstein2014}. For a system with two closely aligned layers, an external magnetic field generates discrete Landau levels (LLs) in which each orbital encloses one flux quantum. We choose the total filling factor $\nu_{\rm tot}$ to be $1$ such that the number of electrons equals the number of Landau orbitals. If one orbital in the upper layer is occupied by an electron, strong interlayer repulsion ensures that the counterpart in the lower layer is vacant. In other words, an electron from the upper layer binds with a hole from the lower layer to become an exciton. By defining the vacuum as the $\nu=1$ integer quantum Hall (IQH) state in the lower layer, the many-body state has an expression that resembles the Bardeen-Cooper-Schrieffer (BCS) state of $s$-wave superconductors. Its properties have been studied extensively in previous theoretical and experimental works~\cite{Fertig1989,Brey1990,Cote1992,ChenXM1992,WenXG1992c,EzawaZF1993,YangK1994,MoonK1995,Spielman2000,Spielman2001,Kellogg2002,Kellogg2004,Tutuc2004,Champagne2008,Tiemann2008a,YoonY2010,Nandi2012}.

Since the electron-hole binding strength depends on the interlayer Coulomb interaction, it is desirable to make interlayer distance as small as possible. For GaAs double quantum wells, there are fundamental constraints due to the width of both wells and the layer separating them. Using graphene and transition metal dichalcogenides (TMDs), van der Waals heterostructures with small interlayer distance can be made. Two classes of double layer samples have been realized in experiments~\cite{LiuXM2017,LiJIA2017b,LiuXM2019,LiJIA2019,LiuXM2022,LinKA2022,ShiQH2022,KimSY2023,LiQX2024}. In both cases, each layer itself could be a monolayer or multilayer graphene and TMDs. For the first class, the two layers are separated by thin layers of hexagonal boron nitride (hBN)~\cite{LiuXM2017,LiJIA2017b,LiuXM2019,LiJIA2019,LinKA2022,LiuXM2022}. For the second class, the two layers are in direct contact but one of them is rotated by a large angle relative to the other~\cite{ShiQH2022,KimSY2023,LiQX2024}. The valleys in their band structures have very different momenta so tunneling between the two layers is strongly suppressed~\cite{Finocchiaro2017}. Exciton condensation at higher temperatures were demonstrated in these samples and new aspects of this phenomenon were explored. For instance, the exciton coupling strength can be tuned in a wide extent to realize the BCS-BEC crossover~\cite{LiuXM2022}. In contrast to GaAs systems, where exciton condensation has only been observed in the lowest LL of GaAs, large-angle twisted samples pave the way for exciton condensation in many other LLs~\cite{ShiQH2022,KimSY2023,LiQX2024}. The nature of low-energy charged excitations may alter with the single-particle eigenstates in LLs~\cite{LiQX2024}.

An exciting recent progress in van der Waals heterostructures is the observation of fractional quantum Hall (FQH) states with excitonic correlations~\cite{ZhangNJ2025,KimDH2025}. It was pointed out that double layer systems at $\nu_{\rm tot}=1/3$ should exhibit features that are similar to the $\nu_{\rm tot}=1$ exciton condensate~\cite{WenXG1992c}, but experimental verification of this prediction remained elusive until the recent breakthrough. In addition, perfect Coulomb drag in the Corbino geometry were observed at several other filling factors and attributed to formation of excitons in FQH states~\cite{ZhangNJ2025}. This paper reports our systematic investigation of exciton physics in the FQH regime. Using the composite fermion theory~\cite{Jain1989a}, we construct a plethora of states that can be understood as composite fermion exciton condensates (CFECs), which include the $\nu_{\rm tot}=1/3$ state as the simplest example. These states can be divided into two types depending on their internal correlations and are realized in different ranges of layer charge imbalance. For certain microscopic models, numerical calculations are performed to corroborate the existence of some states. Experimental signatures of these states are discussed in connection with the phenomena observed in Refs.~\cite{ZhangNJ2025,KimDH2025}. 

\begin{figure}[ht]
\includegraphics[width=0.90\textwidth]{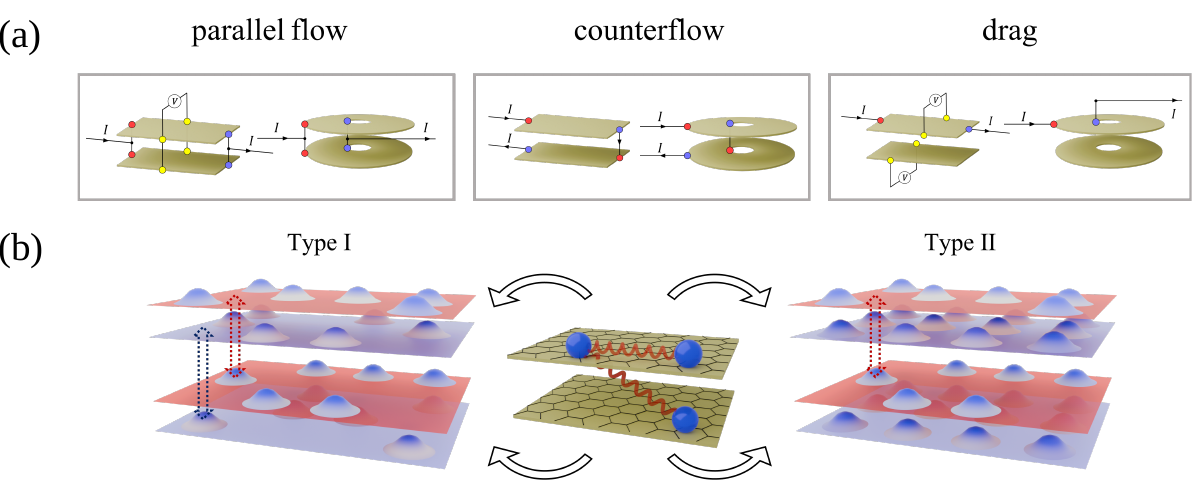}
\caption{(a) Schematics of the sample configurations in parallel flow, counterflow, and drag measurements. (b) Schematics of the double layer system and the two types of electron exciton condensates. If an electron is dressed by two fluxes from each layer, the states would turn into CFECs.}
\label{Figure1}
\end{figure}

\section{Exciton Condensates}
\label{exciton}

\subsection{Electron exciton condensates}

It is helpful to begin with exciton condensation of electrons because their properties can be analyzed more easily. To study excitonic FQH states, we shall map strongly correlated states of electrons to exciton condensate of composite fermions~\cite{Jain1989a}. This not only provides a transparent physical picture but also enables some quantitative calculations. The upper (lower) layer is represented by the symbol $\mathsf{up}$ ($\mathsf{lo}$) in various quantities. For example, the number of electrons in the two layers are denoted as $N_{\mathsf{up}}$ and $N_{\mathsf{lo}}$. In view of the exchange symmetry between the layers, it is sufficient to focus on the cases with $N_{\mathsf{up}} \geq N_{\mathsf{lo}}$. We define complex planar coordinates $\{v_{j}\}$ ($\{w_{j}\}$) for the upper (lower) layer. The most well-known electron exciton condensate is the Halperin 111 wave function~\cite{Halperin1983}
\begin{eqnarray}
\Psi_{111}(\{v_{j}\},\{w_{j}\}) = \prod_{j<k} \left( v_{j} - v_{k} \right) \prod_{j<k} \left( w_{j} - w_{k} \right) \prod_{j,k} \left( v_{j} - w_{k} \right).
\label{WaveFunc111}
\end{eqnarray}
A naive interpretation is that $\Psi_{111}(\{v_{j}\},\{w_{j}\})$ describe double layer systems of non-relativistic electrons confined to the lowest LL with total filling factor $\nu_{\rm tot}=1$. Its utility would be much broader if we turn to the Fock state representation. After choosing a set of single-particle eigenstates, non-interacting basis states $|\Phi_{i}\rangle$ can be defined such that the exciton condensate is defined as a suitable superposition
\begin{eqnarray}
|\Psi_{111}\rangle = \sum_{i} c_{i} |\Phi_{i}\rangle.
\label{FockState111}
\end{eqnarray}
The coefficients $c_{i}$ are chosen such that Eq.~\eqref{WaveFunc111} is recovered when the electrons indeed occupy the non-relativistic lowest LL. If the single-particle eigenstates have other forms (e.g. those in the non-relativistic second LL), we generally do not have simple real space wave functions. By performing electron-hole transformation in the lower layer, the physical picture of Eq.~\eqref{WaveFunc111} becomes transparent. One exciton is formed when an electron in the upper layer binds with a hole in the lower layer and then the excitons form a Bose-Einstein condensate~\cite{Eisenstein2004,Eisenstein2014}. This phase has interesting features related to spontaneous breaking of particle number conservation. In a microscopic model, the two layers are not coupled by tunneling so the number of electrons in both layers are conserved. This can be reformulated as two U(1) symmetries associated with the total number of electrons $N_{e}=N_{\mathsf{up}}+N_{\mathsf{lo}}$ and the pseudospin $S_{z} = (N_{\mathsf{up}} - N_{\mathsf{lo}})/2$. The latter symmetry is spontaneously broken to generate a gapless mode. In other words, the electrons do not possess definite layer index $|\mathsf{up}\rangle$ or $|\mathsf{lo}\rangle$ but choose to enter a coherent superposition $|\mathsf{up}\rangle+\exp(i\phi)|\mathsf{lo}\rangle$. Although Eq.~\eqref{WaveFunc111} can be defined for any $S_{z}$, a highly imbalanced system may realize other phases instead of the exciton condensate~\cite{YangK2001,ZengYH2024,HuZF2024}.

In experimental studies of the exciton condensate, samples are usually prepared in the Hall bar or Corbino geometry and three types of measurements can be performed as shown in Fig.~\ref{Figure1}. In parallel flow measurement, current passes through the two layers along the same direction, and the system behaves as a usual FQH state with quantized Hall resistance and exponentially suppressed longitudinal resistance. In counterflow measurement, current in the two layers have opposite directions, and the system is highly conductive that resembles a non-ideal superfluid~\cite{Kellogg2004,Tutuc2004}. In drag measurement of the Corbino geometry, applying current in one layer induces the same amount of current in the other layer~\cite{Tiemann2008a,Nandi2012}. For our purpose, it is more illuminating to consider drag measurement in the Hall bar geometry. It has been performed in double quantum wells~\cite{Kellogg2002} and van der Waals heterostructures where the two layers are separated by hBN~\cite{LiuXM2017,LiJIA2017b,LiuXM2022}. For large-angle twisted samples, it remains a significant challenge to fabricate individual contacts for the two layers.

When an electric current $I_{\mathsf{up}}$ passes only through the upper layer in a Hall bar sample, a transverse Hall voltage $V_{\mathsf{up}}$ ($V_{\mathsf{lo}}$) emerges in the upper (lower) layer. This allows us to define the upper layer drive Hall resistance $R_{\mathsf{up}}=V_{\mathsf{up}}/I_{\mathsf{up}}$ and the interlayer drag Hall resistance $R_{\rm drag}=V_{\mathsf{lo}}/I_{\mathsf{up}}$. If electric current only flows in the lower layer, its drive Hall resistance $R_{\mathsf{lo}}$ can be obtained. Another drag Hall resistance may be defined in this setting, which should be the same as the previous one based on symmetry analysis. These quantities are organized into a Hall resistance matrix
\begin{eqnarray}
\mathbf{R}_{xy} =
\begin{bmatrix}
R_{\mathsf{up}} & R_{\rm drag} \\
R_{\rm drag} & R_{\mathsf{lo}}
\end{bmatrix}.
\label{HallMatrixGen}
\end{eqnarray}
It was found to be
\begin{eqnarray}
\frac{h}{e^{2}}
\begin{bmatrix}
1 & 1 \\
1 & 1
\end{bmatrix}
\label{HallMatrix111}
\end{eqnarray}
for the Halperin 111 exciton condensate~\cite{Eisenstein2014}, which plays an essential role in later discussions. The derivation of this matrix is not straightforward. In analytical and numerical calculations, the Hall conductance is much more commonly computed than the Hall resistance. When the system is placed on a torus, the Hall conductance is linked to the many-body Chern number defined in the space of twisted boundary conditions~\cite{NiuQ1985,ShengDN2003}. After constructing the Chern-Simons theory of a state, the Hall conductance can be expressed using its $K$ matrix~\cite{WenXG1992a,Renn1992}. If we have accurate trial wave functions for the charged elementary excitations, the Hall conductance can be deduced using the Laughlin flux insertion argument~\cite{Laughlin1981,WuYH2022c}. These methods are not applicable here because Eq.~\eqref{HallMatrix111} is a singular matrix so its inverse (conductance matrix) does not exist. Nevertheless, one can still obtain the result using an argument proposed by Yang~\cite{YangK1998}.

We consider two types of generalizations as illustrated in Fig.~\ref{Figure1}. One common feature of these states is that multiple LLs are populated in the two layers. For the first type, all LLs in the two layers are partially occupied. Electrons in a certain LL of the upper layer binds with holes in the same LL of the lower layer. This certainly seems very strange when multiple LLs in each layer are involved. It is unnatural to leave some orbitals empty in lower LLs but place electrons in higher LLs. However, they turn out to be useful when we study composite fermions. For the second type, the topmost LL in each layer is partially occupied whereas the other LLs are fully occupied. Electrons in the topmost LLs of the two layers form an exciton condensate. This type of states has been observed in van der Waals heterostructures~\cite{LiuXM2017,LiJIA2017b,ShiQH2022,KimSY2023,LiQX2024}. If we neglect mixing between the topmost LL and other levels, all these systems are formally equivalent except that electron-electron interactions depend on the contents of single-particle eigenstates. The first type states can be defined for any $S_{z}$, albeit the competition with other states should be assessed in realisitic systems. In contrast, the second type only exists in the range of $S_{z}$ constrained by reshuffling electrons in the topmost LLs. This is because some electrons in the second type of states do not participate the formation of excitons (those that are not in the topmost LL). 

It is useful to derive the Hall resistance matrix for these states. Let us begin with the first type ones in which each layer has $n_{\rm I}$ partially occupied LLs. The total Hall resistance should be $h/(n_{\rm I}e^{2})$. The voltage-current relation becomes
\begin{eqnarray}
\begin{bmatrix}
V_{\mathsf{up}} \\
V_{\mathsf{lo}}
\end{bmatrix} =
\frac{h}{e^{2}} \frac{1}{n_{\rm I}}
\begin{bmatrix}
1 & 1 \\
1 & 1
\end{bmatrix}
\begin{bmatrix}
I_{\mathsf{up}} \\
I_{\mathsf{lo}}
\end{bmatrix},
\end{eqnarray}
which means that
\begin{eqnarray}
R_{\mathsf{up}} = R_{\mathsf{lo}} = R_{\rm drag} = \frac{1}{n_{\rm I}} \frac{h}{e^{2}}.
\end{eqnarray}
Next we turn to the second type ones in which each layer has $n_{\rm II}$ fully occupied LLs. The current in each layer is decomposed to two parts as $I_{\sigma}=I^{a}_{\sigma}+I^{b}_{\sigma}$ ($\sigma=\mathsf{up},\mathsf{lo}$). The $a$ part is carried by the electrons in the fully occupied LLs whereas the $b$ part is carried by the exciton condensate in the topmost LL. The voltage-current relation for the whole system is
\begin{eqnarray}
\begin{bmatrix}
V_{\mathsf{up}} \\
V_{\mathsf{lo}}
\end{bmatrix} =
\begin{bmatrix}
R_{\mathsf{up}} & R_{\rm drag} \\
R_{\rm drag} & R_{\mathsf{lo}}
\end{bmatrix}
\begin{bmatrix}
I^{a}_{\mathsf{up}} + I^{b}_{\mathsf{up}} \\
I^{a}_{\mathsf{lo}} + I^{b}_{\mathsf{lo}}
\end{bmatrix}
\label{CurrentRelationAll}
\end{eqnarray}
and those for the $a$ and $b$ parts are
\begin{eqnarray}
\begin{bmatrix}
V_{\mathsf{up}} \\
V_{\mathsf{lo}}
\end{bmatrix} =
\frac{h}{e^{2}} 
\begin{bmatrix}
\frac{1}{n_{\rm II}} & 0 \\
0 & \frac{1}{n_{\rm II}} 
\end{bmatrix}
\begin{bmatrix}
I^{a}_{\mathsf{up}} \\
I^{a}_{\mathsf{lo}}
\end{bmatrix}, \qquad 
\begin{bmatrix}
V_{\mathsf{up}} \\
V_{\mathsf{lo}}
\end{bmatrix} =
\frac{h}{e^{2}} 
\begin{bmatrix}
1 & 1 \\
1 & 1
\end{bmatrix}
\begin{bmatrix}
I^{b}_{\mathsf{up}} \\
I^{b}_{\mathsf{lo}}
\end{bmatrix}.
\end{eqnarray}
It is obvious that Eq.~\eqref{CurrentRelationAll} can be converted to
\begin{eqnarray}
\frac{h}{e^{2}}
\begin{bmatrix}
I^{b}_{\mathsf{up}} + I^{b}_{\mathsf{lo}} \\
I^{b}_{\mathsf{up}} + I^{b}_{\mathsf{lo}}
\end{bmatrix} = 
\begin{bmatrix}
R_{\mathsf{up}} & R_{\rm drag} \\
R_{\rm drag} & R_{\mathsf{lo}}
\end{bmatrix}
\begin{bmatrix}
n_{\rm II} \left( I^{b}_{\mathsf{up}} + I^{b}_{\mathsf{lo}} \right) + I^{b}_{\mathsf{up}} \\
n_{\rm II} \left( I^{b}_{\mathsf{up}} + I^{b}_{\mathsf{lo}} \right) + I^{b}_{\mathsf{lo}}
\end{bmatrix}.
\end{eqnarray}
To satisfy this equation for any values of $I^{b}_{\mathsf{up}}$ and $I^{b}_{\mathsf{lo}}$, we must have
\begin{eqnarray}
&& (n_{\rm II}+1) R_{\mathsf{up}} + n_{\rm II} R_{\rm drag} = \frac{h}{e^{2}}, \qquad n_{\rm II} R_{\mathsf{up}} + (n_{\rm II}+1) R_{\rm drag} = \frac{h}{e^{2}}, \\
&& (n_{\rm II}+1) R_{\rm drag} + n_{\rm II} R_{\mathsf{lo}} = \frac{h}{e^{2}}, \qquad n_{\rm II} R_{\rm drag} + (n_{\rm II}+1) R_{\mathsf{lo}} = \frac{h}{e^{2}},
\end{eqnarray}
which are solved to yield
\begin{eqnarray}
R_{\mathsf{up}} = R_{\mathsf{lo}} = R_{\rm drag} = \frac{1}{2n_{\rm II}+1} \frac{h}{e^{2}}.
\end{eqnarray}
In both cases, the Hall resistance matrices are singular.

\subsection{Composite fermion exciton condensates}

As for the Halperin 111 wave function, all subsequent wave functions are written using the lowest LL in the disk geometry, but they should be interpreted as superposition of Fock states that are also useful when other single-particle eigenstates are used. Let us consider the Halperin 333 wave function
\begin{eqnarray}
\Psi_{333}(\{v_{j}\},\{w_{j}\}) = \prod_{j<k} \left( v_{j} - v_{k} \right)^{3} \prod_{j<k} \left( w_{j} - w_{k} \right)^{3} \prod_{j,k} \left( v_{j} - w_{k} \right)^{3}
\label{WaveFunc333}
\end{eqnarray}
at total filling factor $\nu_{\rm tot}=1/3$. Based on a Chern-Simons theory analysis, this state was found to have one gapped sector and one gapless sector~\cite{WenXG1992c}. The former one is the same as the usual Laughlin $1/3$ state in one-component system~\cite{Laughlin1983} and the latter one arises from spontaneous breaking of the U(1) symmetry associated with $S_{z}$. This is an intriguing prediction but its experimental confirmation remained elusive for a long time until the recent experiments~\cite{ZhangNJ2025,KimDH2025}. One primary challenge is that the interlayer distance must be especially small, which is quite difficult to achieve in GaAs yet still possible in van der Waals heterostructures. For our purpose, it is illustrative to rewrite Eq.~\eqref{WaveFunc333} as
\begin{eqnarray}
\Psi_{333}(\{v_{j}\},\{w_{j}\}) = \left[ \prod_{j<k} \left( v_{j} - v_{k} \right) \prod_{j<k} \left( w_{j} - w_{k} \right) \prod_{j,k} \left( v_{j} - w_{k} \right) \right] \prod_{j<k} \left( z_{j} - z_{k} \right)^{2}
\label{WaveFunc333b}
\end{eqnarray}
with $\{z_{i}\}$ being the combination of $\{v_{j}\}$ and $\{w_{j}\}$. This expression can be understood using the composite fermion theory~\cite{Jain1989a}. One electron binds with two fluxes from each layer to become composite fermions as indicated by the factor
\begin{eqnarray}
\prod_{j<k} \left( z_{j} - z_{k} \right)^{2} = \prod_{j<k} \left( v_{j} - v_{k} \right)^{2} \prod_{j<k} \left( w_{j} - w_{k} \right)^{2} \prod_{j,k} \left( v_{j} - w_{k} \right)^{2}.
\end{eqnarray}
The composite fermions experience an effective magnetic field, so they form Landau-like effective levels and realize the Halperin 111 state.

We now replace the Halperin 111 state in Eq.~\eqref{WaveFunc333b} with more general exciton condensates discussed above. This leads to two series of CFECs that are referred to as type-I and type-II. Both series are further divided into two subclasses according to whether the effective magnetic field is parallel or opposite to the actual magnetic field. For the type-I states, the partially filled $n_{\rm I}$ composite fermion levels are described by $\eta^{\rm EC}_{n_{\rm I}}(\{v_{j}\},\{w_{j}\})$. The wave functions for parallel effective magnetic field at $\nu_{\rm tot} = n_{\rm I}/(2n_{\rm I}+1)$ are
\begin{eqnarray}
\Psi^{\rm p}_{n_{\rm I}}(\{v_{j}\},\{w_{j}\}) = \left[ \eta^{\rm EC}_{n_{\rm I}} \right] \prod_{j<k} \left( z_{j} - z_{k} \right)^{2}
\label{TypeOnePos}
\end{eqnarray}
and those for opposite effective magnetic field at $\nu_{\rm tot} = n_{\rm I}/(2n_{\rm I}-1)$ are
\begin{eqnarray}
\Psi^{\rm o}_{n_{\rm I}}(\{v_{j}\},\{w_{j}\}) = \left[ \eta^{\rm EC}_{n_{\rm I}} \right]^{*} \prod_{j<k} \left( z_{j} - z_{k} \right)^{2}.
\label{TypeOneNeg}
\end{eqnarray}
The Halperin 333 state corresponds to $\Psi^{\mathsf{p}}_{n_{\rm I}}(\{v_{j}\},\{w_{j}\})$ with $n_{\rm I}=1$. Although these states can be defined for all $S_{z}$ values, the fate of each member in a specific system is determined by its microscopic details. For the type-II states, each layer has $n_{\rm II}$ fully occupied composite fermion levels and exciton condensate appears in the two topmost effective levels. These two parts are described by the wave functions $\Phi_{n_{\rm II},n_{\rm II}}(\{v_{j}\},\{w_{j}\})$ and $\chi^{\rm EC}_{n_{\rm II}}(\{v_{j}\},\{w_{j}\})$. The wave functions for parallel effective magnetic field at $\nu_{\rm tot} = (2n_{\rm II}+1)/(4n_{\rm II}+3)$ are
\begin{eqnarray}
\Psi^{\rm p}_{n_{\rm II}}(\{v_{j}\},\{w_{j}\}) = \left[ \chi^{\rm EC}_{n_{\rm II}} \otimes \Phi_{n_{\rm II},n_{\rm II}} \right] \prod_{j<k} \left( z_{j} - z_{k} \right)^{2}
\label{TypeTwoPos}
\end{eqnarray}
and those for opposite effective magnetic field at $\nu_{\rm tot} = (2n_{\rm II}+1)/(4n_{\rm II}+1)$ are
\begin{eqnarray}
\Psi^{\rm o}_{n_{\rm II}}(\{v_{j}\},\{w_{j}\}) = \left[ \chi^{\rm EC}_{n_{\rm II}} \otimes \Phi_{n_{\rm II},n_{\rm II}} \right]^{*} \prod_{j<k} \left( z_{j} - z_{k} \right)^{2}.
\label{TypeTwoNeg}
\end{eqnarray}
As for the electron exciton condensate, these states can only be defined for $|S_{z}| \leq \frac{N_{e}}{2(2n_{\rm II}+1)}$, but the actual interval of $S_{z}$ where they are stabilized may be even smaller due to competition with other states. It is evident that two types of states have the same filling factor when $n_{\rm I}=2n_{\rm II}+1$.

We proceed to discuss experimental signatures of these CFECs in electric transports. For parallel flow and counterflow measurements, the results are similar to what happens in electron exciton condensate: the system behaves like a normal FQH state in the former cases and an imperfect superfluid in the latter cases. The Hall responses in drag measurements can be derived using the Chern-Simons mean field theory discussed in Ref.~\cite{LiuXM2019}, albeit in a slightly different context. It provides us the resistivity matrix, which is equivalent to the resistance matrix in two dimensions. This quantity can divided into two parts as 
\begin{eqnarray}
\boldsymbol{\rho}_{xy} =
\begin{bmatrix}
\rho_{\mathsf{up}} & \rho_{\rm drag} \\
\rho_{\rm drag} & \rho_{\mathsf{lo}}
\end{bmatrix} = \boldsymbol{\rho}^{\rm CS}_{xy} \pm \boldsymbol{\rho}^{\rm CF}_{xy}.
\end{eqnarray}
$\boldsymbol{\rho}^{\rm CS}_{xy}$ is the Chern-Simons term solely due to flux attachment, $\boldsymbol{\rho}^{\rm CF}_{xy}$ is the composite fermion term that accounts for their Hall effect, and the plus (minus) sign corresponds to parallel (opposite) effective magnetic field. The former one has the same form
\begin{eqnarray}
\boldsymbol{\rho}^{\rm CS}_{xy} = \frac{h}{e^{2}}
\begin{bmatrix}
2 & 2 \\
2 & 2
\end{bmatrix}
\end{eqnarray}
for all CFECs. The latter one is
\begin{eqnarray}
\boldsymbol{\rho}^{\rm CF}_{xy} = \frac{h}{e^{2}} \frac{1}{n_{\rm I}}
\begin{bmatrix}
1 & 1 \\
1 & 1
\end{bmatrix}
\end{eqnarray}
for type-I states and
\begin{eqnarray}
\boldsymbol{\rho}^{\rm CF}_{xy} = \frac{h}{e^{2}} \frac{1}{2n_{\rm II}+1}
\begin{bmatrix}
1 & 1 \\
1 & 1
\end{bmatrix}
\end{eqnarray}
for type-II states, because the composite fermions are assumed to have the same Hall responses as free electrons. It is also fruitful to attack this problem in a reverse manner. We begin with the assumption that the Hall resistance matrix is singular, which is reasonable as the exciton condensates have gapless sectors. This imposes the constraint $R_{\mathsf{up}}R_{\mathsf{lo}}=R_{\rm drag}R_{\rm drag}$. The exchange symmetry between the two layers indicates that $R_{\mathsf{up}}=R_{\mathsf{lo}}$. In the parallel flow configuration, the total Hall conductance is simply the total filling factor
\begin{eqnarray}
\nu_{\rm tot} \frac{e^{2}}{h} = \frac{1}{R_{\mathsf{up}}+R_{\rm drag}} + \frac{1}{R_{\mathsf{lo}}+R_{\rm drag}}.
\end{eqnarray}
These relations yield the Hall resistance matrix
\begin{eqnarray}
\frac{1}{\nu_{\rm tot}} \frac{h}{e^{2}} \begin{bmatrix}
1 & 1 \\
1 & 1
\end{bmatrix}.
\end{eqnarray}
As a singular matrix, it cannot be derived using other commonly used methods (except the one in Ref.~\cite{YangK1998}). The Hall resistance matrix alone is not able to distinguish between the type-I and type-II states at the same filling factor. An important open question is how to do so in experiements. 

\section{Numerical Results}
\label{numerics}

This section presents numerical results to corroborate the existence of CFECs in microscopic models. We take each layer in the system to be a monolayer graphene (MLG), monolayer TMD, or bilayer graphene (BLG). For all materials, there are two valleys in their band structures and the electrons also carry spin (they are locked together by spin-orbit coupling in TMDs). However, the spin and valley degrees of freedom do not appear in the CFECs. This is not a problem because many-body interaction favor spin and valley polarized states in many cases. The low-energy physics of MLG and TMDs is modeled as massless and massive Dirac fermions, respectively~\cite{XiaoD2012}. The single-particle eigenstates in their LLs are two-component vectors. Since each unit cell in BLG contains four lattice sites, a natural tight-binding description uses one orbital per site~\cite{JungJ2014}. The single-particle eigenstates in its LLs are quite complicated four-component vectors that can only be computed numerically. For our purpose, an approximation can be made to reduce them to two-component vectors (see the Appendix for details). We assume that electrons are confined to one active LL and no mixing with other LLs is allowed. For non-relativistic LLs, the single-particle eigenstates in the $n$-th level are denoted as $|\mathsf{S}_{n}\rangle$. The active Landau orbitals in our system can be written in a unified way as
\begin{eqnarray}
\begin{bmatrix}
g_{0} |\mathsf{S}_{0}\rangle \\
g_{1} |\mathsf{S}_{1}\rangle
\end{bmatrix}.
\label{SingleParticle}
\end{eqnarray}
For the zeroth LL of MLG and TMDs, $g_{1}$ vanishes identically. By tuning the magnetic field in BLG, we can have $g_{1} \approx 0$ in the zeroth LL and $g_{1} \in [0.88,0.95]$ in the first LL.

We place the electrons on a square torus with periodic boundary conditions~\cite{YoshiokaD1983}. Its two sides have lengths $L_{x}$ and $L_{y}$, the total flux through the torus is $N_{\phi}$, and the magnetic length is $\ell_{B}=\sqrt{\hbar/(eB)}$. These variables satisfy the constraint $L_{x}L_{y}=2\pi\ell^{2}_{B}N_{\phi}$. The two layers are separated by a distance $D$. To describe Coulomb interaction, it is convenient to introduce momentum variables $\mathbf{q} = \frac{2\pi}{L_{x}} q_{1} \mathbf{e}_{x} + \frac{2\pi}{L_{y}} q_{2} \mathbf{e}_{y}$ in terms of unit vectors $\mathbf{e}_{x}$ and $\mathbf{e}_{y}$. For two layers labeled as $\sigma$ and $\tau$, the Coulomb potential in momentum space is
\begin{eqnarray}
V_{\sigma\tau}(\mathbf{q}) = \frac{e^{2}}{4\pi\varepsilon\ell_{B}} \frac{2\pi\ell_{B}}{|\mathbf{q}|} \exp\left[ - \left( 1-\delta_{\sigma\tau} \right) D|\mathbf{q}| \right].
\end{eqnarray}
Its coefficient $e^{2}/(4\pi\varepsilon\ell_{B})$ shall be used as the energy scale. We define second quantized operators $C^{\dag}_{\sigma,m}$ and $C_{\sigma,m}$ with $m\in[0,1,\cdots,N_{\phi}-1]$ for the Landau orbitals in each layer. The many-body Hamiltonian is
\begin{eqnarray}
H &=& \frac{1}{2L_{x}L_{y}} \sum_{\sigma,\tau} \sum_{\{m_{i}\}} \sum^{\mathbb{Z}}_{q_{1},q_{2}} V_{\sigma\tau}(\mathbf{q}) \exp \left[ -\frac{1}{2} |\mathbf{q}|^{2} \ell^{2}_{B} - i\frac{2{\pi}q_{1}}{N_{\phi}} \left( m_{1}-m_{4} \right) \right] \; \nonumber \\
&\times& \left[ |g_{0}|^{2} + |g_{1}|^{2} \left( 1 - |\mathbf{q}|^{2}/2 \right) \right]^{2} \; \widetilde{\delta}_{m_{1},m_{3}-q_{2}} \widetilde{\delta}_{m_{2},m_{4}+q_{2}} \; C^{\dag}_{\sigma,m_{1}} C^{\dag}_{\tau,m_{2}} C_{\tau,m_{4}} C_{\sigma,m_{3}},
\label{ManyBodyHamilton}
\end{eqnarray}
where the tilde delta symbol $\widetilde{\delta}_{s,t}$ is one iff $s-t$ is a multiple of $N_{\phi}$. This Hamiltonian can be block diagonalized using center of mass and many-body translation symmetries~\cite{Haldane1985b}. For the filling factor $\nu_{\rm tot}=p/q$, $N_{e}$ and $N_{\phi}$ have the greatest common divisor $N_{\rm gcd}$ that satisfies $N_{e}=pN_{\rm gcd}$ and $N_{\phi}=qN_{\rm gcd}$. There are two conserved momenta $K_{x}$ and $K_{y}$ that take values in $[0,1,\cdots,N_{\rm gcd}-1]$. We only need to examine these symmetry sectors because each eigenvalue repeats itself $q$ times in the full spectra. The definitions can be adjusted carefully such that the ground state occurs at $K_{x}=K_{y}=0$, but this is not needed for our purpose.

\begin{figure}[ht]
\includegraphics[width=0.65\textwidth]{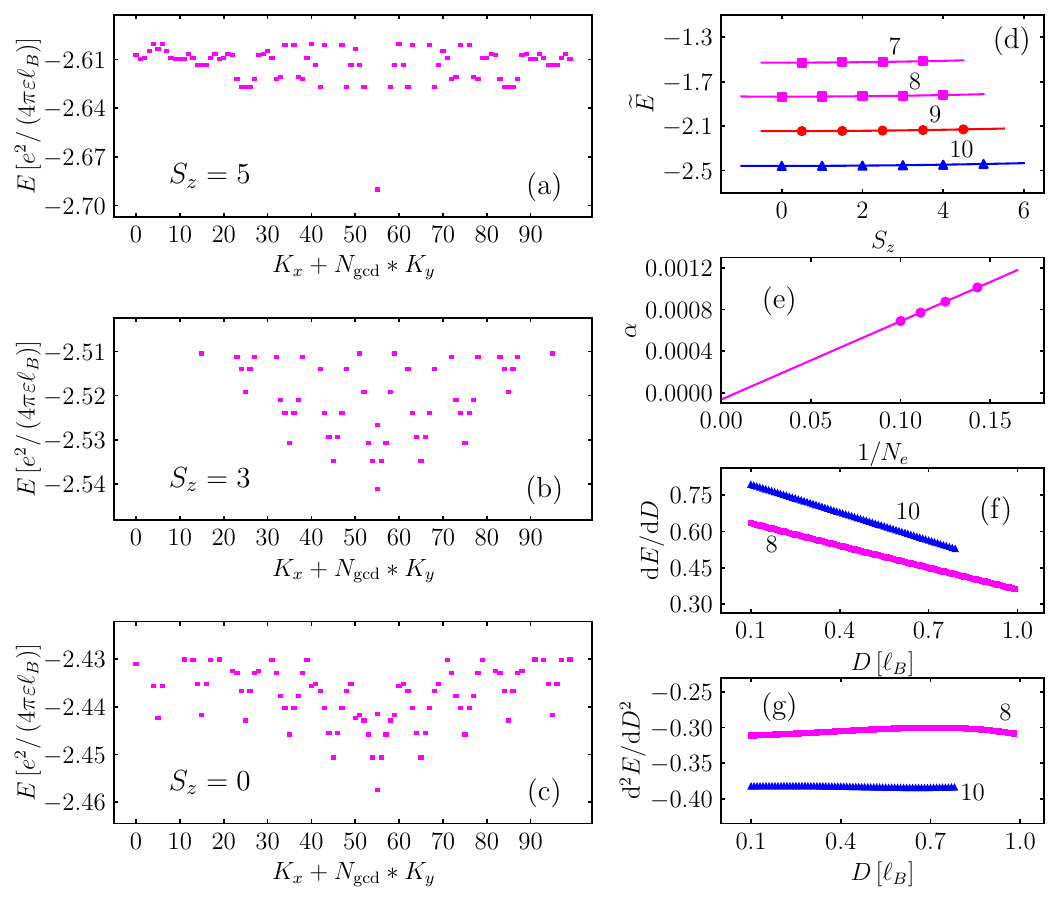}
\caption{Numerical results for type-I CFEC at $\nu_{\rm tot}=1/3$. (a-c) The energy spectra for $N_{e}=10$ at $D=0.3\ell_{B}$. (d) The modified lowest eigenvalues for $N_{e}=7,8,9,10$ at $D=0.3\ell_{B}$ are fitted using Eq.~\eqref{EnergyFormula}. (e) The coefficients $\alpha$ extracted from panel (d) are fitted versus $1/N_{e}$ using a linear function. (e-f) The first and second-order derivatives of the lowest eigenvalues with respect to $D$ for $N_{e}=8,10$ and $S_{z}=0$ in the sector with $K_{x}=K_{y}=\frac{1}{2}N_{e}$. For each curve in panels (d,f,g), $N_{e}$ is indicated by a proximate number.}
\label{Figure2}
\end{figure}

\begin{figure}[ht]
\includegraphics[width=0.65\textwidth]{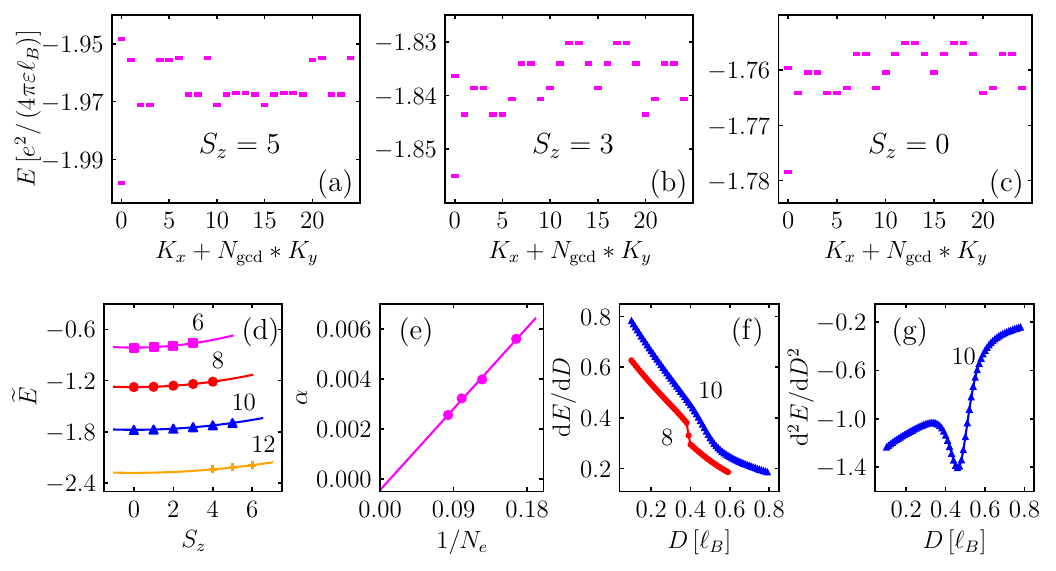}
\caption{Numerical results for type-I CFEC at $\nu_{\rm tot}=2/5$. (a-c) The same data as in Fig.~\ref{Figure2} (a) for $N_{e}=12$. (d) The same data as in Fig.~\ref{Figure2} (d). There are only three points for $N_{e}=12$ since the Hilbert space dimension gets too large at small $S_{z}$. (e) The same data as in Fig.~\ref{Figure2} (e). (f-g) The same data as in Fig.~\ref{Figure2} (f-g) in the sector with $K_{x}=K_{y}=0$.}
\label{Figure3}
\end{figure}

\begin{figure}[ht]
\includegraphics[width=0.65\textwidth]{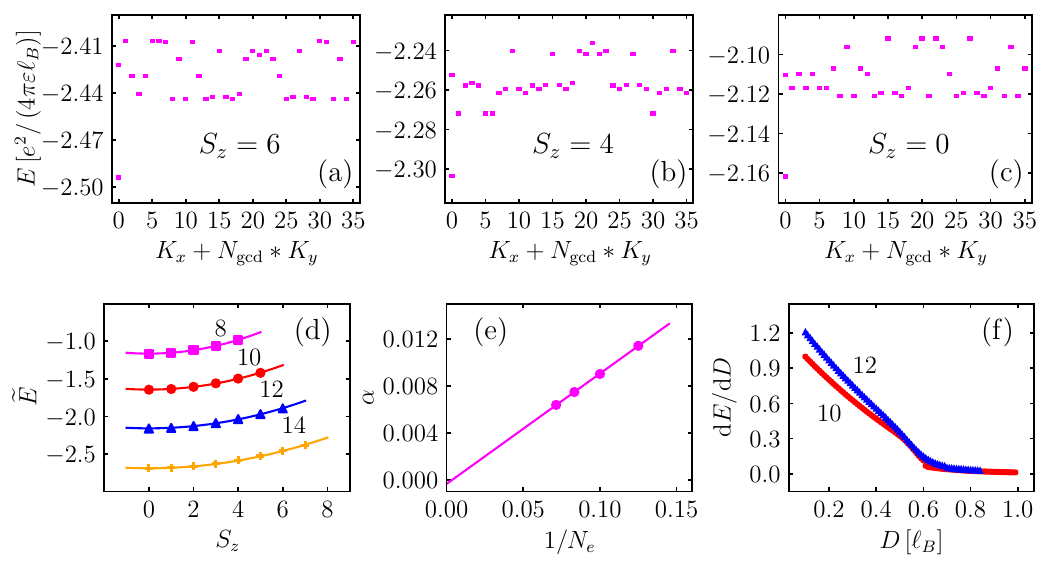}
\caption{Numerical results for type-I CFEC at $\nu_{\rm tot}=2/3$. (a-c) The same data as in Fig.~\ref{Figure2} (a) for $N_{e}=12$. (d) The same data as in Fig.~\ref{Figure2} (d). (e) The same data as in Fig.~\ref{Figure2} (e). (f) The same data as in Fig.~\ref{Figure2} (f) in the sector with $K_{x}=K_{y}=0$.}
\label{Figure4}
\end{figure}

\begin{figure}[ht]
\includegraphics[width=0.65\textwidth]{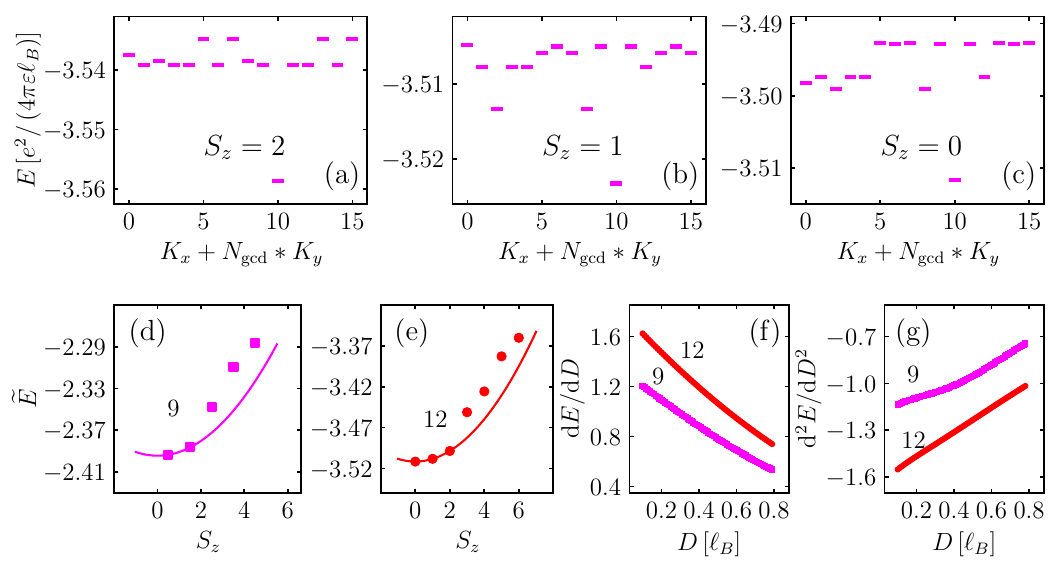}
\caption{Numerical results for type-II CFEC at $\nu_{\rm tot}=3/5$. (a-c) The same data as in Fig.~\ref{Figure2} (a) for $N_{e}=12$. (d-e) The same data as in Fig.~\ref{Figure2} (d). (f-g) The same data as in Fig.~\ref{Figure2} (f-g) in the sector with $K_{x}=K_{y}=0$.}
\label{Figure5}
\end{figure}

Numerical results obtained by exact diagonalization at $\nu_{\rm tot}=1/3,2/5,2/3,3/5$ are presented in Figs.~\ref{Figure2},~\ref{Figure3},~\ref{Figure4},~\ref{Figure5}. The parameter $g_{1}$ in Eq.~\eqref{SingleParticle} is chosen as $0.0$ for $\nu_{\rm tot}=1/3,3/5$ and $0.9$ for $\nu_{\rm tot}=2/5,2/3$. We claim that type-I CFECs are realized at $\nu_{\rm tot}=1/3,2/5,2/3$ and type-II CFEC is realized at $\nu_{\rm tot}=3/5$. To substantiate this conclusion, our first step is to inspect the energy spectra at $D=0.30\ell_{B}$ with $N_{e}$ kept fixed and $S_{z}$ varies. For all cases at $\nu_{\rm tot}=1/3,2/5,2/3$, there is a unique ground state well separated from other excited states at $|S_{z}|=0,1,\cdots,\frac{N_{e}}{2}$ (some examples are displayed in Figs.~\ref{Figure2},~\ref{Figure4},~\ref{Figure5}). The situation is quite different at $\nu_{\rm tot}=3/5$ since this feature is only observed for $|S_{z}|\leq 1$ at $N_{e}=9$ and $|S_{z}|\leq 2$ at $N_{e}=12$ [see Fig.~\ref{Figure3} (a-c)]. After taking into account the center of mass translation, the state at $\nu_{\rm tot}=p/q$ has ground state degeneracy $q$. These results can be explained as follows. When $S_{z}$ reaches the maximal value $N_{e}/2$ at $\nu_{\rm tot}=1/3,2/5,2/3$, all electrons are pushed to the top layer such that they realize the one-component (pseudospin polarized) Jain states. In contrast, the $\nu_{\rm tot}=3/5$ states with $N_{e}=9,S_{z}=1$ and $N_{e}=12,S_{z}=2$ are two-component (pseudospin partially polarized) Jain states. As for the Halperin 111 exciton condensate, we take these states to be the new vacua. To create the states with smaller $S_{z}$, electrons are transferred from the top layer to the bottom layer. This process amounts to the creation of excitons in their respective vacua. It is necessary to begin with different pseudospin values because some electrons in type-II states do not form excitons. 

The next step is to unveil spontaneous symmetry breaking based on quantitative analysis of the lowest eigenvalues and eigenstates in different $S_{z}$ sectors. To compare the eigenvalue, a capacitance energy due to electric charge imbalance must be appended to each value $E$ found in numerical calculations to define the modified value $\widetilde{E}(S_{z})=E+dS^{2}_{z}/N_{\phi}$~\cite{MacDonald1990}. In some cases, a few $S_{z}$ sectors are not available since their Hilbert space dimensions are too large. For the type-I CFECs at $\nu_{\rm tot}=1/3,2/5,2/3$, the results are presented in panels (d) of Figs.~\ref{Figure2},~\ref{Figure3},~\ref{Figure4}, where the data points are fitted by solid lines using the formula
\begin{eqnarray}
\widetilde{E}(S_{z}) = \alpha S^{2}_{z} + \beta.
\label{EnergyFormula}
\end{eqnarray}
As shown in panel (e) of Figs.~\ref{Figure2},~\ref{Figure3},~\ref{Figure4}, the coefficients $\alpha$ exhibit $1/N_{e}$ dependence and decrease to zero as $N_{e}\rightarrow\infty$. For the type-II CFEC at $\nu_{\rm tot}=3/5$, the results for $N_{e}=9$ and $12$ are displayed separately in Fig.~\ref{Figure5} (d) and (e). It is clear that only two and three data points can be fitted using Eq.~\eqref{EnergyFormula} whereas other points deviate from the trend. This further confirms that CFEC is only realized for a certain range of $S_{z}$. Using the coefficient $\alpha$ extracted from $N_{e}=9$ and $12$, linear fitting does not yield zero as $N_{e}\rightarrow\infty$ (no plot is made since there are only two data points). This is somewhat unsatisfactory but probably due to finite size effects. If we do have $\alpha=0$ in the thermodynamic limit for a CFEC, the lowest eigenstates in different $S_{z}$ sectors have the same energy and constitute the ground state subspace. It is possible to construct a new set of eigenstates by putting the electrons in the superposition $|\uparrow\rangle + \exp(i\phi) |\downarrow\rangle$. This is similar to the physics of XY magnet in which spins may point to arbitrary directions in a plane. The phase $\phi$ is uniform in the ground state and its variation in space leads to the excitation encoded by Eq.~\eqref{EnergyFormula}. Finally, the lowest eigenstates $|\Psi(S_{z})\rangle$ in different $S_{z}$ sectors are compared. We transfer one electron between the layers to generate the trial state 
\begin{eqnarray}
|\widetilde{\Psi}(S_{z})\rangle = \sum_{m} C^{\dag}_{\mathsf{lo},m} C_{\mathsf{up},m} |\Psi(S_{z}+1)\rangle.
\label{StateFormula}
\end{eqnarray}
This process builds excitonic correlations into the system. As shown in Table~\ref{Table1}, the overlaps $\langle\widetilde{\Psi}(S_{z})|\Psi(S_{z})\rangle$ are close to $1$ in all cases, so the exact eigenstates indeed represent CFECs. In view of the relations between eigenvalues and eigenstates from two neighboring $S_{z}$ sectors, there should be a zero bias peak in the interlayer tunneling differential conductance. When one electron tunnels from the upper layer to the lower layer, the energy cost is almost zero and the system enters a new state that is basically the same as the ground state at appropriate $S_{z}$. In other words, a large tunneling current would be generated even if an infinitesimal bias is applied. This phenomenon was also predicted and observed in the $\nu_{\rm tot}=1$ electron exciton condensate, but the peak never actually reached infinite~\cite{Eisenstein2014}. Many proposals were made to address this conflict yet no consensus has been arrived~\cite{Balents2001,Stern2001,Fogler2001,Joglekar2001,IwazakiA2003,Fertig2003,Abolfath2004,WangZQ2004,Jack2004,WangZQ2005,Fertig2005,Rossi2005,Klironomos2005,ParkK2006,SuJJ2010,Eastham2010,Hyart2011}.

We now discuss how to interpret the findings reported in Refs.~\cite{ZhangNJ2025,KimDH2025} in light of our numerical calculations. The zeroth LL of MLG was studied in both cases, as captured by our model with $g_{1}=0.0$. In agreement with Refs.~\cite{ZhangNJ2025,KimDH2025}, our results indicate that type-I CFEC was observed at $\nu_{\rm tot}=1/3$. There are also signatures of excitons at $\nu_{\rm tot}=2/5,2/3,3/5$~\cite{ZhangNJ2025}, but their origins require more careful examinations. An important feature is that perfect interlayer drag only exist in a limited range where $\nu_{\mathsf{up}} \approx \nu_{\rm tot}$ and $\nu_{\mathsf{lo}}$ is small. In other words, we may begin with completely imbalanced systems with all electrons in the upper layer and then transfer a few portion of electrons to the lower layer. Excitonic correlations develop between the layers initially and disappear after passing certain thresholds. This behavior is incompatible with the type-I CFECs studied here since they persist down to $\nu_{\mathsf{up}}=\nu_{\mathsf{lo}}$. To resolve this discrepancy, we have performed additional calculations at $\nu_{\rm tot}=2/5,2/3,3/5$ using $g_{1}=0.0$. Instead of the CFECs, balanced system realizes a two-component Halperin state at $\nu_{\mathsf{up}}=\nu_{\mathsf{lo}}=1/5$ and a Jain state at $\nu_{\mathsf{up}}=\nu_{\mathsf{lo}}=1/3$~\cite{Halperin1983,WuXG1993}. The physics at small $\nu_{\mathsf{lo}}$ is more intricate. While one-component FQH states are realized at $S_{z}=\frac{N_{e}}{2}$, the energy spectra at $S_{z}=\frac{N_{e}}{2}-1$ depend sensitively on system parameters. For Coulomb interaction and $D=0.3\ell_{B}$, the lowest eigenstates at $S_{z}=\frac{N_{e}}{2}-1$ and $\frac{N_{e}}{2}$ do not have the same momenta in some cases, so they are not related to each other via Eq.~\eqref{StateFormula}. Motivated by the results on LL mixing in MLG~\cite{Peterson2013}, we add zeroth and first Haldane pseudopotentials~\cite{Haldane1983c} to the many-body Hamiltonian (with coefficients $-0.21$ and $-0.05$, respectively). In such cases, excitonic correlations are consistently found between the lowest eigenstates at $S_{z}=\frac{N_{e}}{2}$ and $\frac{N_{e}}{2}-1$: operating on the former ones according to Eq.~\eqref{StateFormula} generates good approximations for the latter ones. It is unfortunately not possible to estimate the interval of $\nu_{\mathsf{lo}}$ where type-I CFECs can be realized due to limited system sizes. We note that both types of CFECs can be found at $\nu_{\rm tot}=3/5$ in different ranges of $S_{z}$. At intermediate $S_{z}$, the states may not belong to either type. In particular, there is a two-component Jain state at $\nu_{\mathsf{up}}=2/5,\nu_{\mathsf{lo}}=1/5$ with Hall resistance matrix
\begin{eqnarray}
\frac{h}{e^{2}} \begin{bmatrix}
3/2 & 2 \\
2 & 1
\end{bmatrix}.
\end{eqnarray}
By tuning $\nu_{\mathsf{lo}}$ across $1/5$ under appropriate conditions, switch between two different Hall responses can be observed at $\nu_{\rm tot}=3/5$. 

\begin{table}
\begin{tabular}{cc|cccccc}
\hline
\hline
                &         &        &       & $S_{z}$ &        &        &        \\
                          \cline{3-8}          
$\nu_{\rm tot}$ & $N_{e}$ &    5   &    4   &    3   &    2   &    1   &   0    \\
\hline
1/3             &    10   &   ---  & 1.0000 & 0.9993 & 0.9993 & 0.9994 & 0.9996 \\
2/5             &    10   &   ---  & 0.9932 & 0.9789 & 0.9753 & 0.9704 & 0.9759 \\
2/3             &    12   & 1.0000 & 0.9728 & 0.9737 & 0.9759 & 0.9833 & 0.9855 \\
3/5             &    12   &   ---  &   ---  &   ---  &  ---   & 0.9841 & 0.9900 \\
\hline
\hline
\end{tabular}
\caption{The overlaps of exact eigenstates $|\Psi(S_{z})\rangle$ and trial states $|\widetilde{\Psi}(S_{z})\rangle$ in Eq.~\eqref{StateFormula}. Absence of states in some cases are indicated by lines.}
\label{Table1}
\end{table}

\section{Phase Transitions}
\label{transit}

We have chosen the interlayer distance $D$ to be small because the CFECs are only stabilized in such cases. It is possible to construct other candidates for a given $\nu_{\rm tot}$ and phase transitions between different states are expected when $D$ is tuned. This problem has been studied extensively in the electron exciton condensate at $\nu_{\rm tot}=1$. The decoupled limit $D\rightarrow\infty$ is relatively simple. In the balanced case with $\nu_{\mathsf{up}}=\nu_{\mathsf{lo}}=1/2$, each layer may realize a composite fermion liquid (CFL) or Moore-Read type FQH states~\cite{Halperin1993,Moore1991,Levin2007,LeeSS2007,SonDT2015,Zucker2016}. In the unbalanced case with suitable $\nu_{\mathsf{up}}$ and $\nu_{\mathsf{lo}}$, two FQH states that are particle-hole conjugates of each other could be found. Numerical calculations suggest that some of these states are separated from the exciton condensate by continuous phase transitions~\cite{ChenH2012,ZhuZ2019,ZhangYH2023}. Many theoretical ideas about other phases at intermediate $D$ have been discussed but they are not yet confirmed in experiments~\cite{KimYB2001,Veillette2002,Stern2002,Simon2003,ParkK2004,Moller2008,Moller2009,Alicea2009,Milovan2015,IsobeH2017,Sodemann2017,Potter2017,ZhuZ2017b,LianB2018a,Wagner2021,Ruegg2023,DengHY2024,Lotric2024,Ruegg2024}. An exhaustive survey of all possible states in our models is too demanding. We shall only consider balanced cases for simplicity. The decoupled limit at large $D$ is relatively easy to analyze. For the cases with $\nu_{\rm tot}=2/5,2/3$, simple one-component states are expected at $\nu=1/5,1/3$. Depending on sample details, Wigner crystal, CFL, or non-Abelian FQH state may appear at $\nu=1/6$~\cite{WangCY2025,Balram2025}. There is no simple state at $\nu=3/10$ but paired CFs may generate a non-Abelian FQH state~\cite{Mukherjee2015}. 

To understand the physics at intermediate $D$, we adopt a powerful principle gleaned from the composite fermion theory. For an electron in one specific layer, the number of fluxes attached to it from the other layer is dictated by the strength of interlayer repulsion. Its value is $2$ for the CFECs at small $D$ and should decreases to $1$ after $D$ passes certain thresholds. Indeed, previous works have identified a series of states with two (one) intralayer (interlayer) attached fluxes~\cite{LiuXM2019,LiJIA2019,Faugno2020a}. Based on this principle, we conjecture that the states at intermediate $D$ are 
\begin{eqnarray}
&& \Psi_{\frac{1}{3}}(\{v_{j}\},\{w_{j}\}) = \left[ \prod_{j<k} \left( v_{j} - v_{k} \right) \prod_{j<k} \left( w_{j} - w_{k} \right) \right] \prod_{j<k} \left( v_{j} - v_{k} \right)^{4} \prod_{j<k} \left( w_{j} - w_{k} \right)^{4} \prod_{j,k} \left( v_{j} - w_{k} \right), \\
&& \Psi_{\frac{2}{5}}(\{v_{j}\},\{w_{j}\}) = \left[ \Phi_{\rm FS}(\{v_{j}\}) \Phi_{\rm FS}(\{w_{j}\}) \right] \prod_{j<k} \left( v_{j} - v_{k} \right)^{4} \prod_{j<k} \left( w_{j} - w_{k} \right)^{4} \prod_{j,k} \left( v_{j} - w_{k} \right), \\
&& \Psi_{\frac{2}{3}}(\{v_{j}\},\{w_{j}\}) = \left[ \Phi_{\rm FS}(\{v_{j}\}) \Phi_{\rm FS}(\{w_{j}\}) \right] \prod_{j<k} \left( v_{j} - v_{k} \right)^{2} \prod_{j<k} \left( w_{j} - w_{k} \right)^{2} \prod_{j,k} \left( v_{j} - w_{k} \right), \\
&& \Psi_{\frac{3}{5}}(\{v_{j}\},\{w_{j}\}) = \left[ \Phi_{3}(\{v_{j}\}) \Phi_{3}(\{w_{j}\}) \right] \prod_{j<k} \left( v_{j} - v_{k} \right)^{2} \prod_{j<k} \left( w_{j} - w_{k} \right)^{2} \prod_{j,k} \left( v_{j} - w_{k} \right).
\end{eqnarray}
For the first (last) two states, the intralayer attached fluxes is $4$ ($2$). The composite fermions realize $\nu=1$ IQH state, Fermi sea state $\Phi_{\rm FS}$, and $\nu=3$ IQH state $\Phi_{\rm 3}(\{v_{j}\})$, respectively. We thus conclude that the first and fourth ones are gapped FQH states whereas the second and third ones are gapless CFLs. Using the $K$ matrix formalism, the first (fourth) state is found to have ground state degeneracy $24$ ($40$) on the torus~\cite{WenXG1992a}. For system sizes that can be reached in exact diagonalization, it is unlikely that these large numbers can be discerned. The CFLs are also difficult to identify in numerical calculations because they depend sensitively on the system size and shape of the torus. In particular, low-energy states may shift among momentum sectors as $N_{e}$ changes. For double MLG [$g_{1}=0.0$ in Eq.~\eqref{SingleParticle}], the $\nu_{\rm tot}=2/3$ CFL was found in a small range of $D$~\cite{Faugno2020a}. There are also many other candidates at $\nu_{\rm tot}=2/3$~\cite{Geraedts2015,Peterson2015,LiuZ2015a}. If the Fermi seas in $\Psi_{\frac{2}{5}}(\{v_{j}\},\{w_{j}\})$ and $\Psi_{\frac{2}{3}}(\{v_{j}\},\{w_{j}\})$ are replaced by paired CFs, the whole system would become gapped non-Abelian FQH states, but we expect that specially designed interactions are need to stabilize them. To probe phase transitions between the CFECs and the intermediate states, first and second-order derivatives of the ground state energy have been computed [see Fig.~\ref{Figure2} (f-g),~\ref{Figure3} (f-g),~\ref{Figure4} (f),~\ref{Figure5} (f-g)]. Both derivatives are continuous at $\nu_{\rm tot}=1/3,3/5$ up to the largest available $D$. This is quite appealing as the system may exhibit higher-order phase transitions. For $\nu_{\rm tot}=2/3$ with $N_{e}=10,12$, the first-order derivatives are discontinuous. The results at $\nu_{\rm tot}=2/5$ are somewhat puzzling: the first-order derivative is discontinuous for $N_{e}=8$ and continuous for $N_{e}=10$, and the second-order derivative for $N_{e}=10$ has a dip. An obvious explanation for the discontinuities is that first-order transitions occur, but they may also be attributed to peculiar features of the CFLs. The derivatives for $N_{e}=10$ at $\nu_{\rm tot}=2/5$ imply that a continuous transition is still possible. One should be cautious about these claims due to the limited system sizes and the uncertainty about intermediate states. 

\section{Conclusions}

In conclusion, we have constructed two series of CFECs and demonstrate some of them can be found in realistic microscopic models. Experimental signatures of these states include superfluid like counterflow transport, perfect interlayer Coulomb drag, and zero-bias tunneling peak. The states recently observed in experiments should be type-I CFECs according to our analysis~\cite{ZhangNJ2025,KimDH2025}. An important difference is that the CFEC at $\nu_{\rm tot}=1/3$ is stable in the whole range of $\nu_{\mathsf{lo}}$ whereas those at $\nu_{\rm tot}=2/5,2/3,3/5$ only appear if $\nu_{\mathsf{lo}}$ is small enough. Using the first LL of BLG, type-I CFECs should be found at $\nu_{\rm tot}=2/5,2/3$ in wider ranges of $\nu_{\mathsf{lo}}$. It is further unveiled that a type-II CFEC can be realized at $\nu_{\rm tot}=3/5$ when the two layers are nearly balanced ($|\nu_{\mathsf{up}}-\nu_{\mathsf{lo}}|$ is small). In principle, two types of CFECs may also coexist at other filling factors such as $\nu_{\rm tot}=3/7$, but numerical calculations are too diffcult. By tuning charge densities of the two layers, type-II CFECs may turn into some intermediate states and then to type-I CFECs. This leads to sudden alterations in the Hall conductance matrix. We have also discussed other competing states in balanced systems. The results are still preliminary but quantum phase transitions with exotic properties can be anticipated when the interlayer distance is tuned. For two types of CFECs at the same filling factor, the Hall conductance matrix cannot tell them apart. To address this problem, more investigations about the CFECs in connection with other experimental techniques should be carried out. Besides the tunneling between two layers, previous works have discussed sideways tunneling between two exciton condensates at $\nu_{\rm tot}=1,1/m$~\cite{WenXG1996,WangTL2024}. The possibility that excitons form crystalline states has been studied~\cite{YangK2001,ZengYH2024,HuZF2024}. Similar problems can be explored in more general CFECs and we hope that interesting results will be obtained.

\section*{Acknowledgments}

Some calculations were performed using the DiagHam package for which we are grateful to the authors~\cite{DiagHam}. X. J. H. and Y. H. W. were supported by National Natural Science Foundation of China (Grant No. 12174130). L. W. was supported by the National Key Projects for Research and Development of China (Grant Nos. 2021YFA1400400, 2022YFA1204700) and Natural Science Foundation of Jiangsu Province (Grant No. BK20220066).

\setcounter{figure}{0}
\setcounter{table}{0}
\setcounter{equation}{0}
\renewcommand{\thefigure}{A\arabic{figure}}
\renewcommand{\thetable}{A\arabic{table}}
\renewcommand{\theequation}{A\arabic{equation}}


\appendix

\section{Landau levels in bilayer graphene}

This appendix reviews the tight-binding model and LLs of BLG. We use one orbital per lattice site to construct the tight-binding model. The distance between Carbon atoms is $a=0.142$ nm. There are five Slonczewski-Weiss-McClure hopping parameters $\gamma_{i}$ ($i=0,1,\cdots,4$). Another parameter $\delta$ is introduced to characterize the energy difference between dimer and non-dimer lattice site. Based on ab initio calculations, their values were found to be $\gamma_{0}=2.61$, $\gamma_{1}=0.361$, $\gamma_{2}=0.0$, $\gamma_{3}=0.283$, $\gamma_{4}=0.138$, $\gamma_{5}=0.0$, and $\delta=0.015$ (in units of eV)~\cite{JungJ2014}. If a perpendicular electric field is applied, the potential energy in the two layers are denoted as $\Delta$ and $-\Delta$. At the corners of the Brillouin zone, there are two valleys $\mathbf{K}_{\pm}$. In the vicinity of $\mathbf{K}_{+}$ point, momentum values are described by their deviations $k_{x}$ and $k_{y}$ from $\mathbf{K}_{+}$. The tight-binding model can be expanded to yield
\begin{eqnarray}
\mathcal{H}_{\mathbf{K}_{+}} =
\begin{bmatrix}
\Delta & \frac{3}{2}a\gamma_{0}k_{-} & -\frac{3}{2}a\gamma_{4}k_{-} & -\frac{3}{2}a\gamma_{3}k_{+} \\
\frac{3}{2}a\gamma_{0}k_{+} & \Delta+\delta & \gamma_{1} & -\frac{3}{2}a\gamma_{4}k_{-} \\
-\frac{3}{2}a\gamma_{4}k_{+} & \gamma_{1} & -\Delta+\delta & \frac{3}{2}a\gamma_{0}k_{-} \\
-\frac{3}{2}a\gamma_{3}k_{-} & -\frac{3}{2}a\gamma_{4}k_{+} & \frac{3}{2}a\gamma_{0}k_{+} & -\Delta
\end{bmatrix}
\end{eqnarray}
with $k_{\pm}=k_{x} \pm ik_{y}$. Next a magnetic field generated by vector potential $\mathbf{A}$ is applied along the vertical direction. This requires minimal coupling substitution in the Hamiltonian
\begin{eqnarray}
\hbar k_{\pm} \rightarrow \Pi_{\pm} = \hbar k_{\pm} - e \left( A_{x} \pm iA_{y} \right).
\end{eqnarray}
As in non-relativistic Landau problem, we define the ladder operators
\begin{eqnarray}
\widehat{\mathsf{a}} = \frac{\ell_{B}}{\sqrt{2}\hbar} \Pi_{-}, \qquad \widehat{\mathsf{a}}^{\dag} = \frac{\ell_{B}}{\sqrt{2}\hbar} \Pi_{+}
\end{eqnarray}
such that the number operator $\widehat{\mathsf{a}}^{\dag}\widehat{\mathsf{a}}$ has eigenstates $|\mathsf{S}_{n}\rangle$. This leads to the Hamiltonian
\begin{eqnarray}
\begin{bmatrix}
\Delta & \frac{3a\gamma_{0}}{\sqrt{2}\ell_{B}} \widehat{\mathsf{a}} & -\frac{3a\gamma_{4}}{\sqrt{2}\ell_{B}} \widehat{\mathsf{a}} & -\frac{3a\gamma_{3}}{\sqrt{2}\ell_{B}} \widehat{\mathsf{a}}^{\dag} \\
\frac{3a\gamma_{0}}{\sqrt{2}\ell_{B}} \widehat{\mathsf{a}}^{\dag} & \Delta+\delta & \gamma_{1} & -\frac{3a\gamma_{4}}{\sqrt{2}\ell_{B}} \widehat{\mathsf{a}} \\
-\frac{3a\gamma_{4}}{\sqrt{2}\ell_{B}} \widehat{\mathsf{a}}^{\dag} & \gamma_{1} & -\Delta+\delta & \frac{3a\gamma_{0}}{\sqrt{2}\ell_{B}} \widehat{\mathsf{a}} \\
-\frac{3a\gamma_{3}}{\sqrt{2}\ell_{B}} \widehat{\mathsf{a}} & -\frac{3a\gamma_{4}}{\sqrt{2}\ell_{B}} \widehat{\mathsf{a}}^{\dag} & \frac{3a\gamma_{0}}{\sqrt{2}\ell_{B}} \widehat{\mathsf{a}}^{\dag} & -\Delta
\end{bmatrix},
\end{eqnarray}
whose eigenstates can be written generally as
\begin{eqnarray}
\begin{bmatrix}
f_{00} |\mathsf{S}_{0}\rangle + f_{01} |\mathsf{S}_{1}\rangle + f_{02} |\mathsf{S}_{2}\rangle + \ldots \\
f_{10} |\mathsf{S}_{0}\rangle + f_{11} |\mathsf{S}_{1}\rangle + f_{12} |\mathsf{S}_{2}\rangle + \ldots \\
f_{20} |\mathsf{S}_{0}\rangle + f_{21} |\mathsf{S}_{1}\rangle + f_{22} |\mathsf{S}_{2}\rangle + \ldots \\
f_{30} |\mathsf{S}_{0}\rangle + f_{31} |\mathsf{S}_{1}\rangle + f_{32} |\mathsf{S}_{2}\rangle + \ldots
\end{bmatrix}.
\end{eqnarray}
Since each row is an infinite summation, the coefficients $f_{in}$ can only be computed numerically using proper truncation. For the zeroth and first LLs, the single-particle eigenstates can be approximated by
\begin{eqnarray}
|{\rm LL}_{0}\rangle = \begin{bmatrix}
0 \\
0 \\
0 \\
|\mathsf{S}_{0}\rangle 
\end{bmatrix} \qquad
|{\rm LL}_{1}\rangle = \zeta \begin{bmatrix}
0 \\
f_{10} |\mathsf{S}_{0}\rangle \\
f_{20} |\mathsf{S}_{0}\rangle \\
f_{31} |\mathsf{S}_{1}\rangle 
\end{bmatrix}
\end{eqnarray}
because other coefficients are quite small ($\zeta$ is chosen to normalize the vector). If they are used to construct the many-body Hamiltonian, the results would have the same form as Eq.~\eqref{ManyBodyHamilton} with $g_{0}=1$ for $|{\rm LL}_{0}\rangle$ and $|g_{0}|^{2}=|\zeta f_{10}|^{2}+|\zeta f_{20}|^{2},|g_{1}|^{2}=|\zeta f_{31}|^{2}$ for $|{\rm LL}_{1}\rangle$. The physics in the $\mathbf{K}_{-}$ valley is similar.

\bibliography{../ReferConde}

\end{document}